\documentclass[aps,amssymb,amsmath,twocolumn, pra]{revtex4}
\usepackage{graphicx}
\usepackage{bm}

\usepackage{color}
\usepackage{hyperref}
\usepackage[up]{subfigure}

\newcommand{\be}{\begin{equation}}
\newcommand{\ee}{\end{equation}}
\newcommand{\bc}{\begin{center}}
\newcommand{\ec}{\end{center}}
\newcommand{\bea}{\begin{eqnarray}}
\newcommand{\eea}{\end{eqnarray}}
\newcommand{\ra}{\rangle}
\newcommand{\la}{\langle}

\begin{document}
\title{Entanglement dynamics under local Lindblad evolution}
\author{Sandeep K. \surname{Goyal}}
\email{goyal@imsc.res.in}
\affiliation{The Institute of mathematical Sciences, CIT campus, Chennai 600 113, India}
\author{Sibasish \surname{Ghosh}}
\email{sibasish@imsc.res.in}
\affiliation{The Institute of mathematical Sciences, CIT campus, Chennai 600 113, India}
\begin{abstract}
The phenomenon of entanglement sudden death (ESD) in finite dimensional
composite open systems is described here for both bi-partite as well
as multipartite cases, where  individual subsystems  undergo
Lindblad type heat bath evolution. ESD is found to be generic for non-zero
temperature of the bath. At $T=0$, one-sided action of the
heat bath on pure entangled states of two qubits does not show ESD.
\end{abstract}

\maketitle





\section{Introduction}
Entanglement is considered to be the most useful resource in Quantum
Information Theory \cite{Nielsen}: it is essential for
quantum teleportation, superdense coding, communication complexity
problem, one-way computation etc.
 Not only that creating entangled
state is a non-trivial task, to store or transmit entangled states in
an error-free manner is also difficult, if not impossible, due to the
very fragile character of quantum system-- every quantum system has a
high possibility to interact with its environment, and thereby, the
system will, in general, get entangled with its environment. This will
give rise to the phenomenon of decoherence \cite{Zurek}.

In general, the purity of any initial state of the quantum system goes
down with time in the presence of decoherence. This decoherence  time
depends on the system as well as on the character of the interaction of
the system with its environment. So, due to the monogamy property of
entanglement, the initial entanglement (if any) of a bipartite or
multipartite quantum system will, in general, decay (to zero) when each
individual system undergoes a decoherence procedure. What can be said
about the associated rate of the above-mentioned decay in entanglement?
How does one compare the rate of decoherence of the individual
subsystems and the rate of decay in initial entanglement among the
subsystems? In this connection, Yu and Eberly  \cite{YuEbr} described a
phenomenon called ``entanglement sudden death (ESD)'' in which the
entanglement decay rate is shown to be exponentially larger than the
rate of decoherence. This happens  whenever the individual qubits of a two-qubit
system undergo evolution under local heat bath action at zero
temperature.

ESD was shown for a certain class of two-qubit states which were initially entangled
 at zero temperature ($T=0$). It was then generalized for $T\ne
0$  for ``$X$'' states \cite{AsDa}. In this paper, we want to find out
the largest set of two-qubit states each of which undergoes ESD due
to  Markovian local heat bath dynamics. If the initial
bath state is the thermal state, at finite temperature ($T\ne 0$) it
is shown that all two-qubit states show ESD. The same is true also where
initial states of the bath is a squeezed thermal state. On the other hand, under
 one sided  Markovian heat bath evolution (i.e, only
one of the two qubits is undergoing Markovian heat bath evolution),
one does not see any ESD at $T=0$ for any two-qubit pure state if the
initial state of the bath is the vacuum state. 
However, we will find ESD if the initial state of the bath is the
squeezed vacuum. This feature does
not hold for one-sided  {\em quantum non demolition} (QND)
evolution  \cite{Banerjee08} 
with squeezed thermal state as the initial state of the bath. In this case, no
two-qubit pure entangled state show ESD.
In deriving all this, we have used the factorization rule for concurrence
\cite{Konrad08}. We have also
described the ESD phenomena for states of $d\otimes 2$ systems as well
as those of $n$-qubits by using the above factorization rule.

The structure of the paper is as follows:
in section \ref{2qubitcase}, we  discuss the ESD in two-qubit
system with the bath acting only on one qubit. In section
\ref{dx2case}, we  generalize the factorization law for
entanglement decay \cite{Konrad08}  to $d\otimes 2$ dimensional
systems.  In section \ref{n-qubit}, we  use the result of
section \ref{dx2case} and show ESD in $n$-qubit systems. In section
\ref{squeezedsec}, we  show ESD when the bath
is in squeezed thermal state initially. In section \ref{QND}, we 
consider the QND-type evolution of the individual qubits. Finally  in section \ref{dxd}
we will derive the sufficient condition for ESD in all
dimensions. Section \ref{conc}  contains our conclusion and
discussions on some open problems.
\section{Two-qubit under local bath}\label{2qubitcase}
We study a $2$-qubit system initially prepared in an 
entangled state where one of them (say the first one) is interacting  with a
reservoir at temperature $T$. 
If the initial state of the bath in contact of the system under
consideration be the thermal state, then the dynamics of a
single-qubit density matrix $\rho$ describing the system (under the
Born-Markov-rotating wave approximation) is given by: 
\begin{align}
\frac{d\rho}{dt} =& \frac{(N+1)\gamma}{2}\left[
  2\sigma_-\rho\sigma_+
  -\sigma_+\sigma_-\rho-\rho\sigma_+\sigma_-\right]\nonumber\\  
&+ \frac{(N)\gamma}{2}\left[
  2\sigma_+\rho\sigma_-
  -\sigma_-\sigma_+\rho-\rho\sigma_-\sigma_+\right] \label{mastereq}
\end{align}
where
$N$ is the mean occupation number of the reservoir, $\gamma$ 
is the spontaneous decay rate of the 
qubits, $\sigma_+ = |1\ra\la 0|$ and $\sigma_- = |0\ra\la
1|$. Here we are ignoring the unitary part of the evolution which is
irrelevant for our purpose \cite{AsDa, Breuer}. We can rewrite eq$^n$(\ref{mastereq}) as:
\begin{align}
\dot{\rho} &= \Lambda[\rho]\Rightarrow \dot{\rho}_{ij} = \sum_{kl}L_{ij,kl}\rho_{kl}\label{supop}
\end{align}
where $L$ is the  matrix representation for $\Lambda$ (called Lindblad
operator \cite{lindblad76}). 
The solution for  eq$^n$(\ref{supop}) is:
\begin{align}
\rho(t)_{ij} &= \sum_{kl}V_{ij,kl}\rho(0)_{kl}\label{supop02}.
\end{align}
Here $V=\exp(Lt)$ \cite{ACH07}
is  a completely  positive map as $\rho(t)$
is a valid density matrix of the qubit at any time $t$. Our aim is to find
entanglement in the evolved $2$-qubit state after applying the map $V$
on one qubit. For that purpose we will use the {\em factorization law
  for  entanglement decay} 
\cite{Konrad08} which states that the concurrence of a two qubit  pure
state $|\chi\rangle$-- with one qubit being subject to an arbitrary
channel $V$-- will satisfy the equation $C(\rho_{AB}) =
C\left(|\chi\rangle\right)C\left((\mathbb{I}\otimes 
V)(|\phi^+\rangle\langle \phi^+|)\right)$ where $\rho_{AB}=
(\mathbb{I}\otimes V) \left(|\chi \rangle\langle \chi|\right)$. To show
{\em entanglement sudden death} (ESD) in any state \cite{YuEbr}, it is sufficient
to show that the state $|\phi^+\rangle$ evolves to a separable state
under the action of given single qubit operation $V$. We can calculate the $V$ matrix
for our case:
\begin{align}
\label{evo-ch}
V &= \left[\begin{array}{cccc}
\frac{1+x^2}{2} + x\cot(\theta) & 0 & 0 & y_2\\
0& x & 0&0\\
0&0& x &0\\
y_1 & 0&0& \frac{1+x^2}{2}-x\cot(\theta)
\end{array}\right]
\end{align}
where $y_1 = 2x\cot(\theta)N,
~~y_2 = 2x\cot(\theta)(1+N),~~
\tan(\theta) = \left(\frac{2x(1+2N)}{x^2-1}\right),~~
x = \exp\left[-\frac{1}{2}\gamma(1+2N)t\right]$. When we apply this
$V$ on one side of $|\phi^+\rangle$ we will get the mixed state:
\begin{align}
M(t) &= \left[\begin{array}{cccc}
\frac{1+x^2}{2} + x\cot(\theta) & 0 & 0 & x \\
0& y_2& 0&0\\
0&0&y_1&0\\
 x  & 0&0& \frac{1+x^2}{2}-x\cot(\theta)
\end{array}\right]
\end{align}
For the matrix $M(t)$ to represent a  separable state, the partial transpose of
$M(t)$ should be a positive semi-definite matrix \cite{Asher96,
  Horodecki96}. If $\sinh(\gamma(1+2N)t \ge 
2\sqrt{\frac{N(N+1)}{(1+2N)^2} }$, the matrix $M(t)$ is positive
semi-definite under
partial transposition and hence is separable. 
This shows  that  the operator $V$ acting on one-qubit evolves the maximally
entangled state $|\phi^+\rangle$ into a separable state.
The factorization law for entanglement decay implies that
all the two qubit pure entangled states show ESD.

Now consider the zero temperature case ($T=0$).
For $T=0$ the mean
occupation number $N$ is zero and hence $y_1$ is also zero. The matrix
$M$ matrix will never
be positive under partial transposition and hence will always
represent an entangled state. The direct implication of this result is
that there is no pure two qubit entangled state that will show ESD at
$T=0$.

For $T\ne 0$, every two-qubit pure state shows ESD and since mixed
states are convex combinations of pure states, every two-qubit mixed
state will show ESD. However, at $T=0$ some mixed states do show ESD
even though no pure state does.

In section \ref{n-qubit} we will extend our result to
the $n$-qubit case. But before that we will derive the generalized
factorization law for entanglement decay, i.e, the factorization law
for $d\otimes 2$ systems in the next section.
\section{Factorization law for entanglement decay for $d\otimes 2$
  systems}\label{dx2case}

\begin{center}
\begin{figure}
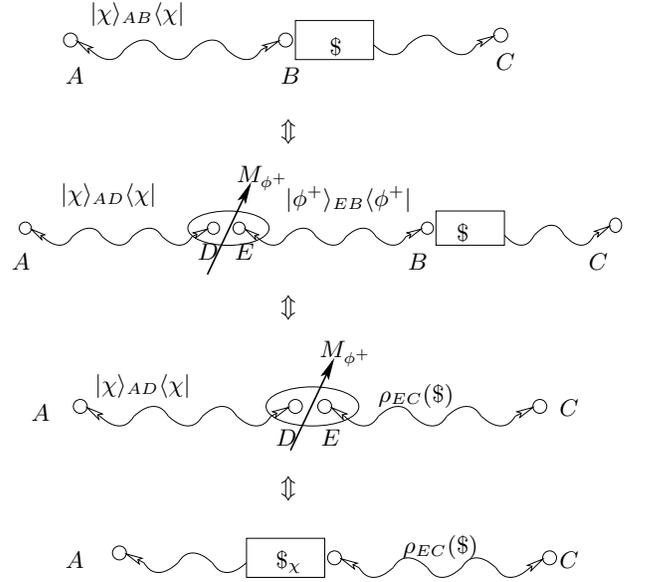
\caption{Duality between the pairs ($|\chi\rangle\langle\chi|,~\$$)
  and ($\$_{\chi},~\rho(\$)$) via Choi-Jamiolkowski isomorphism.} 
  \label{fig1} 
\end{figure}
\end{center}

In this section, we consider the effect of any single-qubit
trace-preserving map $\$ $ on side $B$ of any bipartite density matrix
$\rho_{AB}$, where dim$\mathcal{H}_A = d$. In ref. \cite{Li09} one can
see some discussion on the factorization law for entanglement decay
for $d\otimes d$ systems. 

To start with, we consider first its effect on any pure state
$|\chi\rangle_{AB}$ having Schmidt form $|\chi\rangle_{AB} =
\sqrt{p}|00\rangle_{AB}+\sqrt{1-p}|11\rangle_{AB}$. Here
$|0\rangle_B,~|1\rangle_B$ are the eigenvectors of $\sigma_z$ for the
subsystem $B$ and $|0\rangle_A,~|1\rangle_A$ are taken from the
standard orthonormal basis $\{
|0\rangle_A,~|1\rangle_A,\cdots,|d-1\rangle_A\}$ of $\mathcal{H}_A$
with $0<p<1$.

As $\left(\mathbb{I}_A\otimes
\$\right)\left(|\chi\rangle_{AB}\langle\chi|\right)$ is a two-qubit
density matrix, one can find out its concurrence
$\mathcal{C}(\left(\mathbb{I}_A\otimes
\$\right)\left(|\chi\rangle_{AB}\langle\chi|\right))$. So, according to
fig.[\ref{fig1}], we have 
\begin{align}
&  \mathcal{C}(\left(\mathbb{I}_A\otimes
  \$\right)\left(|\chi\rangle_{AB}\langle\chi|\right))
  =\nonumber\\
&\mathcal{C}\left(\mathcal{N}: \langle \phi^+|_{DE}\left(
  |\chi\rangle_{AD}\langle\chi|\otimes\rho_{EC}(\$)\right)|\phi^+\rangle_{DE}:\right),
\end{align}
where $\rho_{EC}= \left(\mathbb{I}_E\otimes
\$\right)\left(|\phi^+\rangle_{EC}\langle\phi^+|\right),~\mathcal{N}:Z:
\equiv Z/({\rm Tr}Z)$ and $|\phi^+\rangle 
= \frac{1}{\sqrt{2}}(|00\rangle+|11\rangle)$ is a
two qubit maximally entangled state.

We now perform the POVM $\{\mathcal{M} = \sqrt{p}|0\rangle_A\langle 0| +
\sqrt{1-p}|1\rangle_a\langle 1|,~ \mathbb{I}_{d\times d}-\mathcal{M}\}$
on side $A$ of the $d\otimes 2$ density matrix $\rho_{AC}(\$)$.So in
the case when $\mathcal{M}$ is clicked, the 
normalized output state  is
given by
\begin{align}
\mathcal{N}: \left(\mathcal{M}\otimes
\mathcal{I}_C\right)\rho_{AC}(\$)\left(\mathcal{M}\otimes
\mathcal{I}_C\right):&\equiv \sigma_{AC} \mbox{ say}.
\end{align}
Note here that $\rho_{AC}(\$)$ is same as $\rho_{EC}(\$)$ with just
$E$ replaced by $A$. Along the line of \cite{Konrad08}, it will follow
immediately that 
\begin{align}
\mathcal{N}: \langle \phi^+|_{DE}\left(
  |\chi\rangle_{AD}\langle\chi|\otimes\rho_{EC}(\$)\right)|\phi^+\rangle_{DE}:
  &= \sigma_{AC}.
\end{align}
Thus we see that $\left(\mathbb{I}_A\otimes
\$\right)\left(|\chi\rangle_{AB}\langle\chi|\right) = \sigma_{AC}$. So $\mathcal{C}(\left(\mathbb{I}_A\otimes
\$\right)\left(|\chi\rangle_{AB}\langle\chi|\right)) =
\mathcal{C}(\sigma_{AC})$. Now the analysis of \cite{Konrad08} will
follow straight away and we will get:
\begin{align}
\mathcal{C}(\left(\mathbb{I}_A\otimes
\$\right)\left(|\chi\rangle_{AB}\langle\chi|\right)) =&
\mathcal{C}\left((\mathbb{I}_A\otimes\$)(|\phi^+\rangle_{AB}\langle\phi^+|)\right)\nonumber\\
&\times \mathcal{C}(|\chi\rangle_{AB}\langle\chi|).
\end{align}
Next we consider the case when $|\chi\rangle_{AB}$ has the following
Schmidt decomposition:
\begin{align}
|\chi\rangle_{AB}&=
(U_A\otimes U_B)(\sqrt{p}|00\rangle_{AB}+\sqrt{1-p}|11\rangle_{AB})\nonumber\\
&=(U_A\otimes U_B)|\chi_0\rangle_{AB}
\end{align}
where $U_A$ is a $d\times d$ unitary matrix while $U_B$ is a $2\times
2$ unitary matrix.

\begin{widetext}
\begin{align}
(I_A \otimes \$)(|{\chi}{\rangle}_{AB}{\langle}{\chi}|) = {\cal N} :
    ({\cal M}^{\prime} \otimes I_C)\{(I_A \otimes
    \$)(|{\phi}_1^+{\rangle}_{AB}{\langle}{\phi}_1^+|)\}({\cal
      M}^{\prime} \otimes I_C) : \equiv {\sigma}_{AC} \mbox{ (say) },
\end{align}
 where
    ${\cal M}^{\prime} = U_A{\cal M}U_A^{\dagger} =
    U_A({\sqrt{p}}|0{\rangle}_A{\langle}0| + {\sqrt{1 -
        p}}|1{\rangle}_A{\langle}1|)U_A^{\dagger}$ and
    $|{\phi}_1^+{\rangle}_{AB} = (U_A \otimes
    U_B)|{\phi}^+{\rangle}_{AB}$.  
One can then write 
\begin{align}
{\sigma}_{AC} = {\cal N} : (U_A \otimes
I_B)[({\cal M} \otimes I_C)\{(I_A \otimes \$)((I_A \otimes
  U_B)|{\phi}^+{\rangle}_{AB}{\langle}{\phi}^+|(I_A \otimes
  U_B^{\dagger}))\}({\cal M} \otimes I_C)](U_A^{\dagger} \otimes I_B):
\end{align}
\end{widetext}
 Now $(I_A \otimes U_B)|{\phi}^+{\rangle}_{AB} = (V_A \otimes
I_B)|{\phi}^+{\rangle}_{AB}$ where the $2 \times 2$ unitary matrix
$U_B$ is acting on the two dimensional space spanned by $\{|0{\rangle}_B,
|1{\rangle}_B\}$ and the $d \times d$ unitary matrix $V_A$ acts on the
linear span of $\{|0{\rangle}_A, |1{\rangle}_A\}$ and is extended
unitarily on the entire space spanned by $\{|0{\rangle}_A,
|1{\rangle}_A, \ldots, |d - 1{\rangle}_A\}$. This means that both
$V_A|0{\rangle}_A$ as well as $V_A|1{\rangle}_A$ lie inside the linear
span of $\{|0{\rangle}_A, |1{\rangle}_A\}$. Note that the same is true
for $U_B$ also.  

So, here ${\sigma}_{AC} =  {\cal N} : (U_A \otimes I_B)[({\cal M}V_A
  \otimes I_C)\{(I_A \otimes
  \$)(|{\phi}^+{\rangle}_{AB}{\langle}{\phi}^+|)\}(V_A^{\dagger}{\cal
    M} \otimes I_C)](U_A^{\dagger} \otimes I_B) : \equiv {\cal N} :
(U_A \otimes I_B) {\tau}_{AC}(U_A^{\dagger} \otimes I_B) :$  

Therefore, ${\cal C}({\sigma}_{AC}) = {\cal C} ({\tau}_{AC})$.   

Now 
\begin{align}
&{\tau}_{AC}{\tilde{\tau}}_{AC} = {\tau}_{AC}({\sigma}_y \otimes
{\sigma}_y){\tau}_{AC}^*({\sigma}_y \otimes {\sigma}_y)\nonumber\\
& = ({\cal M}V_A
\otimes I_C){\rho}_{AC}(\$)(V_A^{\dagger}{\cal M} \otimes
I_C)({\sigma}_y \otimes {\sigma}_y)\nonumber\\
&\times({\cal M}V_A^* \otimes
I_C){\rho}_{AC}(\$)^*(V_A^T{\cal M} \otimes I_C)({\sigma}_y \otimes
{\sigma}_y),
\end{align}
 which will, in turn, show that ${\rm det}
[{\tau}_{AC}{\tilde{\tau}}_{AC} - {\lambda}I_{4 \times 4}] =~ {\rm
  det} [{\eta}_{AC}({\cal M}{\sigma}_y{\cal M} \otimes
  {\sigma}_y){\eta}_{AC}^*({\cal M}{\sigma}_y{\cal M} \otimes
  {\sigma}_y) - {\lambda}I_{4 \times 4}]$ with ${\eta}_{AC} = (V_A
\otimes I_C){\rho}_{AC}(\$)(V_A^{\dagger} \otimes I_C)$.  

Thus the concurrence of $\tau_{AC}$ is same as that of the two-qubit
state $\eta_{EC}$. So $\mathcal{C}(\left(\mathbb{I}_A\otimes
\$\right)\left(|\chi\rangle_{AB}\langle\chi|\right)) =
\mathcal{C}(\eta_{EC})$. Now, for the state $\eta_{EC}$, we have
already seen that the factorization rule holds good for
concurrence. Therefore, we have  for any pure state
$|\chi\rangle$ of a $d\otimes 2$ system:
\begin{align}
&\mathcal{C}(\left(\mathbb{I}_A\otimes
\$\right)\left(|\chi\rangle_{AB}\langle\chi|\right)) \nonumber\\
=&\mathcal{C}\left((\mathbb{I}_A\otimes\$)(|\phi^+\rangle_{AB}\langle\phi^+|)\right)
 \mathcal{C}(|\chi_0\rangle_{AB}\langle\chi_0|)\nonumber\\
=&
\mathcal{C}\left((\mathbb{I}_A\otimes\$)(|\phi^+\rangle_{AB}\langle\phi^+|)\right)
\mathcal{C}(|\chi\rangle_{AB}\langle\chi|)
\end{align}

Let $\rho_{AB}$ be a mixed state on $\mathbb{C}^d\otimes
\mathbb{C}^2$. Let us consider an arbitrary ensemble representation
for $\rho_{AB}$:
\begin{align}
\rho_{AB}& = \sum_{j=1}^Np_j|\phi_i\rangle_{AB}\langle\phi_j|, \mbox{
  with }\\
E_F(\rho_{AB})&\le \sum_{j=1}^Np_jE_F(|\phi_j\rangle_{AB}\langle\phi_j|),
\end{align}
$E_F(\rho)$ being the entanglement of formation. Now
\begin{align}
E_F(|\phi_j\rangle_{AB}\langle\phi_j|)& = f\left(\mathcal{C}(|\phi_j\rangle_{AB}\langle\phi_j|)\right),
\end{align}
where $f(x)$ is a monotonic function of $x\in [0,1]$. Here
\begin{align}
&E_F((\mathbb{I}_A\otimes \$)(\rho_{AB}))\nonumber \\
&= E_F\left(\sum_{j=1}^Np_j(\mathbb{I}_A\otimes
\$)(|\phi_j\rangle_{AB}\langle\phi_j|)\right)\nonumber \\
&\le \sum_{j=1}^Np_jE_F\left((\mathbb{I}_A\otimes
\$)(|\phi_j\rangle_{AB}\langle\phi_j|)\right)\nonumber\\
&=\sum_{j=1}^Np_jf\left(\mathcal{C}\left((\mathbb{I}_A\otimes
\$)(|\phi_j\rangle_{AB}\langle\phi_j|)\right)\right)\nonumber\\
&= \sum_{j=1}^Np_jf\left(\mathcal{C}\left((\mathbb{I}_A\otimes
\$)(|\phi^+\rangle_{AB}\langle\phi^+|)\right)\right)
\mathcal{C}\left(|\phi_j\rangle_{AB}\langle\phi_j|\right)  
\end{align}
Nevertheless, when $\mathcal{C}\left((\mathbb{I}_A\otimes
\$)(|\phi^+\rangle_{AB}\langle\phi^+|)\right)$ becomes zero
$f\left(\mathcal{C}\left((\mathbb{I}_A\otimes
\$)(|\phi^+\rangle_{AB}\langle\phi^+|)\right)
\mathcal{C}\left(|\phi_j\rangle_{AB}\langle\phi_j|\right)\right)$ will
automatically  become
zero. So, in that case,  $E_F((\mathbb{I}_A\otimes\$)(\rho_{AB})) =
0$.

\section{ESD in $n$-qubit system}\label{n-qubit}
In this section we will be dealing with the evolution of entanglement
of an $n$-qubit system. We will prove that at finite temperature, all the
$n$-qubit states show ESD under the map $V$-- given in
eq$^n$(\ref{evo-ch})-- acting on each qubit 
locally. To make the proof less cumbersome we will do it for the three-qubit
case only. Let the state of the three-qubit system be $|\psi\rangle_{ABC}$
evolving under the bath $V\otimes V\otimes V$. The final state is:
\begin{align}
\rho & = (V\otimes V\otimes V)\left(|\psi\rangle_{ABC}\langle\phi|\right)\nonumber\\
&= (\mathbb{I}\otimes\mathbb{I}\otimes V) (\mathbb{I}\otimes
V\otimes\mathbb{I})(V\otimes\mathbb{I}\otimes\mathbb{I})
\left(|\psi\rangle_{ABC}\langle\phi|\right)\nonumber\\  
&= (\mathbb{I}\otimes\mathbb{I}\otimes V) (\mathbb{I}\otimes
V\otimes\mathbb{I})\rho_{A:BC}\nonumber\\
&= (\mathbb{I}\otimes\mathbb{I}\otimes V) \rho_{A:B:C}
\end{align}
where $\rho_{A:BC}= (V\otimes\mathbb{I}\otimes\mathbb{I})
\left(|\psi\rangle_{ABC}\langle\phi|\right)$ and $\rho_{A:B:C}= (\mathbb{I}\otimes
V\otimes\mathbb{I})\rho_{A:BC}$. We have seen in section
\ref{2qubitcase} that $V$ evolves $|\phi^+\rangle$ to a separable
state in some time $t$, and so the factorization law for entanglement decay
for $2\otimes d$ system implies that the state $\rho_{A:BC}$ will be 
separable in the partition $A:BC$ at time $t$. As this time $t$ is the
maximum time any state 
can take to lose entire entanglement by the action of channel $V$ on
one subsystem, we can see that at the same time $t$ the reduced state
$\rho_{BC}\equiv {\rm Tr}[\rho_{A:BC}]$ becomes separable in the partition $B:C$. Hence we get
the full separability in the partition $A:B:C$. So all the three-qubit pure
states show ESD. Mixed states are the convex sum of pure states. If
all the pure states show ESD, all the mixed states will also show
ESD. This result can be generalized to
any the $n$-qubit case and  we will get full $n$ separability at time $t$.
\section{ Squeezed thermal bath}\label{squeezedsec}
In this section we will consider the case of an $n$-qubit system $S$
interacting with a squeezed thermal bath acting locally on each of the
$n$ individual qubits. The evolution of the reduced
density matrix of the system $S$ in the interaction picture has the
form \cite{Banerjee08, Scully, Breuer}:
\begin{align}\label{squeezedbath}
&\frac{d}{dt}\rho^s(t) =\nonumber\\
&\gamma_0(N+1)
\left(\sigma_-\rho^s(t)\sigma_+-\frac{1}{2}\sigma_+\sigma_-\rho^s(t)
-\frac{1}{2}\rho^s(t)\sigma_+\sigma_-\right)\nonumber\\
&+\gamma_0N\left(\sigma_+\rho^s(t)\sigma_--\frac{1}{2}
\sigma_-\sigma_+\rho^s(t)
-\frac{1}{2}\rho^s(t)\sigma_-\sigma_+\right)\nonumber\\
&-\gamma_0M\sigma_+\rho^s(t)\sigma_+-\gamma_0M^*\sigma_-\rho^s(t)\sigma_-
\end{align}
\noindent Here $\gamma_0$ is the spontaneous emission rate given by
$\gamma_0=4\omega^2|{\bf d}|^2/3\hbar c^3$, and $\sigma_+,~\sigma_-$
are the standard raising and lowering operators, respectively given
by
$\sigma_+ = |1\rangle\langle 0|; ~~\sigma_- = |0\rangle\langle 1|$ and
$2N +1  = \cosh(2r)(2N_{th}+1),~~
M= -\frac{1}{2}\sinh(2r)e^{i\phi}(2N_{th}+1),~~
N_{th}= (e^{\frac{\bar{h}\omega}{k_BT}}-1)^{-1}$.
Here $N_{th}$ is  Planck's distribution giving the number of thermal
photons at the frequency $\omega$,  $r$ and $\phi$ are squeezing
parameters, ${\bf d}$ is the transition matrix elements of the
dipole operator and $c$ is the speed of light in vacuum. 
We have neglected the
Hamiltonian evolution part in
eq$^n$(\ref{squeezedbath}) as we did in the evolution equation
(\ref{mastereq}). Moreover, 
eq$^n$(\ref{squeezedbath}) is in a Lindblad form and hence it
corresponds to a completely positive map $V$ \cite{Breuer, Scully}.

From  equation(\ref{squeezedbath})  we can find our
$V_{sq}$ matrix which can be written as:
\begin{align}
V_{sq} &= \left[\begin{array}{cccc}
\alpha &0&0&\beta\\
0&ye^{-i\omega t}&ze^{-i\omega t}&0\\
0&z^*e^{i\omega t}&ye^{i\omega t}&0\\
\mu &0&0&\nu\end{array}\right],
\end{align}
where
\begin{align}
\alpha &= \frac{N(1+x^2)+x^2}{2N+1}\nonumber\\
\beta &= \frac{N(1-x^2)}{2N+1}\nonumber\\
\mu &= \frac{(N+1)(1-x^2)}{2N+1}\nonumber\\
\nu &= \frac{N(1+x^2) +1}{2N+1}\nonumber\\
x^2 &=  \exp[-\gamma_0(2N+1)t]\nonumber\\
y &= \cosh\left(\frac{\gamma_0at}{2}\right)x\nonumber\\
z &= \sinh\left(\frac{\gamma_0at}{2}\right)x \exp\left[i\Phi
\right]\nonumber\\
a &= \sinh(2r)(2N_{th}+1)\nonumber
\end{align}
The state corresponding to this channel $V_{sq}$ can be given by
Choi-Jamiolkowski isomorphism \cite{jamiolkowski, choi75}. The corresponding state
is:
\begin{align}
(\mathbb{I}\otimes V_{sq} )(|\phi^+\rangle\langle \phi^+|) &=
M_{sq}\nonumber\\
&=\left[\begin{array}{cccc}
\alpha &0&0&ye^{-i\omega t}\\
0&\beta&ze^{-i\omega t}&0\\
0&z^*e^{i\omega t}&\mu &0\\
ye^{i\omega t}&0&0&\nu\end{array}\right]
\end{align}
This matrix $M_{sq}$ is a positive semi-definite matrix.
The positivity under partial transposition, i.e,
\begin{align}
\alpha\nu - |z|^2 \ge 0,~~
\beta\mu - y^2\ge 0,\label{ppt1}
\end{align} equivalently 
\begin{align}
\frac{N(N+1)}{2N+1}4\cosh^2\left(\frac{\gamma_0(2N+1)t}{2}\right)
-\sinh^2\left(\frac{\gamma_0at}{2}\right)&\ge 0,\label{ppt2}\\
\frac{N(N+1)}{2N+1}4\sinh^2\left(\frac{\gamma_0(2N+1)t}{2}\right)
-\cosh^2\left(\frac{\gamma_0at}{2}\right)&\ge 0.\label{ppt3}
\end{align} ensures separability.
It is enough to show that  equation (\ref{ppt3}) is true in order to show
that the state $M_{sq}$ is separable as
$\cosh^2\left(\frac{\gamma_0(2N+1)t}{2}\right) \ge
\sinh^2\left(\frac{\gamma_0(2N+1)t}{2}\right)$ and 
$\sinh^2\left(\frac{\gamma_0at}{2}\right)\le 
\cosh^2\left(\frac{\gamma_0at}{2}\right)$. At $t=0$ this condition (\ref{ppt3}) is
violated. At very large $t $, this condition is true if $2N+1 > a$, i.e,
$2N+1 > \sinh(2r)(2N_{th}+1)$. Now $\sinh(2r)\le \cosh(2r)$ which
implies $2N+1
> \sinh(2r)(2N_{th}+1)$, since $2N+1 = \cosh(2r)(2N_{th}+1)$. This shows
that there exists a finite time $t$ at which $M_{sq}$ will turn from
an entangled
state to a separable state. This shows that one sided operation of $V_{sq}$
on $|\phi^+\rangle\langle\phi^+|$ evolves it into a separable state. Now
using the factorization law for entangled states for $d\otimes 2$ case
(as was done in section \ref{n-qubit}), we can say that all $n$-qubit states shows ESD.

When $r=0$, i.e, when there is no squeezing then $\sinh(2r)=0$ and $2N +1$ is always
greater than $\sinh(2r)(2N_{th}+1)$. This shows that when there is
squeezing, the system takes more time to lose the entanglement. So
instead of choosing the thermal bath one can choose squeezed thermal
bath and delay the ESD process.

At $T=0$, ESD does not occur for a pure two-qubit entangled state 
if one considers the effect of $V$ on one of the qubits where the initial
state of the bath is a vacuum and the evolution is governed by
equation(\ref{mastereq}). But for the case of squeezed bath, at
$T=0$, the mean occupation number $N$ is not zero and hence all the
pure state show ESD. {\em This implies that at zero temperature one
  can reduce the time of dissipation of the entanglement by switching
  on squeezing. Conversely, switching on squeezing at  $T\ne 0$ delays ESD.}

\section{Qubit in quantum non demolition (QND) interaction with bath}\label{QND}
In this section we will study the evolution of a system of qubits in
QND interaction with bath. QND open quantum systems are those systems
in which the system Hamiltonian commutes with the interaction
Hamiltonian. In this section we will be considering the evolution
which
Banerjee et al. has considered in \cite{Banerjee08}. We can write the
CP map $V_{QND}$ and the $M_{QND}$ following 
 equation (10) of reference \cite{Banerjee08}:
\begin{align}
V_{QND}=\left(\begin{array}{cccc}
1&0&0&0\\
0&e^{-i\omega t}e^{-(\hbar\omega)^2\gamma(t)}&0&0\\
0&0&e^{i\omega t}e^{-(\hbar\omega)^2\gamma(t)}&0\\
0&0&0&1
\end{array}\right),\\
M_{QND}=\left(\begin{array}{cccc}
1&0&0&e^{-i\omega t}e^{-(\hbar\omega)^2\gamma(t)}\\
0&0&0&0\\
0&0&0&0\\
e^{i\omega t}e^{-(\hbar\omega)^2\gamma(t)}&0&0&1
\end{array}\right),
\end{align}
where $\omega$ is the natural frequency of the system and 
$\gamma(t)$ is the time dependent spontaneous decay parameter
(see \cite{Banerjee08}). We can see that the matrix $M_{QND}$ is not
positive under partial transposition and hence under the one-sided
action of $V_{QND}$, $|\phi^+\rangle\langle\phi^+|$ will not become
separable. Therefore this map will not show ESD.
\section{Sufficient condition for ESD in any finite dimensional system}\label{dxd}
In this section we will derive the sufficient condition for ESD for
any multipartite system where the dimension of each subsystem is
$d$. For that purpose we will consider the state-channel duality for
$d\otimes d$ system. Consider a map $\$ $ acting on the $B$ subsystem,
where the state of the system $AB$ is
$|\chi \rangle_{AB} = \sum_{i = 1}^d\sqrt{p_i}|ii\rangle$. The action of the channel on the state
$|\chi\rangle_{AB}$ can be
written as:
\begin{align}
&\left(\mathbb{I}_A\otimes
  \$\right)\left(|\chi\rangle_{AB}\langle\chi|\right)
  =\nonumber\\
&\mathcal{N}: \langle \Phi^+|_{DE}\left(
  |\chi\rangle_{AD}\langle\chi|\otimes\rho_{EC}(\$)\right)|\phi^+\rangle_{DE}:\nonumber\\
 =&
  \mathcal{N}:(\mathcal{M}_A\otimes\mathbb{I}_C)\rho_{AC}
  (\mathcal{M}_A\otimes\mathbb{I}_C)^{\dagger}: \nonumber\\
\Rightarrow &E(\left(\mathbb{I}_A\otimes
  \$\right)\left(|\chi\rangle_{AB}\langle\chi|\right))= \nonumber\\
 & E(  \mathcal{N}:(\mathcal{M}_A\otimes\mathbb{I}_C)\rho_{AC}
  (\mathcal{M}_A\otimes\mathbb{I}_C)^{\dagger}:).
\end{align}
where $\mathcal{M}_A = \sum p_i|i\rangle\langle i|$ and $\rho_{AC} =
(\mathbb{I}\otimes \$)(|\phi^+\rangle\langle\phi^+|)$.
If $\$ $ turns out to be an entanglement breaking channel
\cite{Horodecki03}  (equivalently,
if $(\mathbb{I}_A\otimes \$)(|\phi^+\rangle_{AB}\langle\phi^+|)$ is
separable), then $\mathcal{N}:(\mathcal{M}_A\otimes\mathbb{I}_C)\rho_{AC}
  (\mathcal{M}_A\otimes\mathbb{I}_C)^{\dagger}:$ is also
separable and hence, $E(\left(\mathbb{I}_A\otimes
  \$\right)\left(|\chi\rangle_{AB}\langle\chi|\right)) =0$. Thus
  separability of  $\left(\mathbb{I}_A\otimes
  \$\right)\left(|\phi^+\rangle_{AB}\langle\phi^+|\right)$, for a given
  $\$ $, is sufficient for ESD. 

Now, based on the discussion in  section \ref{dx2case}, we can claim
  that if we have a channel $\$ $ which shows ESD in $d\otimes d$
  systems, then the same operation, acting locally on each subsystem,
  will show ESD in multipartite   system as well.

\section{Conclusion}\label{conc}
In this paper we have shown that if the initial
bath state is the thermal state, at finite temperature ($T\ne 0$)
all two-qubit states as well as multiqubit states show ESD. The same
result has been shown for squeezed thermal initial states of the bath. 
We have found that under
the one-sided action of the Markovian heat bath evolution, 
one does not see any ESD at $T=0$ in two-qubit pure state where the
initial state of the bath is the vacuum state. This result is in  sharp
contrast with the case when the initial state of the bath is squeezed
vacuum in which all two-qubit states show ESD.
It has been shown that squeezing delays ESD in the finite temperature case
but speeds it up at zero temperature. 
We have found that for the  action of one-sided  QND evolution  
with squeezed thermal state as the initial state of the bath, no
two-qubit pure entangled state show ESD. Finally we have derived
the sufficient conditions for the ESD for any multipartite system of
identical (but distinguishable) particles. There are articles where
one can find ESD in the systems evolving under non-Markovian evolution
\cite{Ikram07}. 

After having the full understanding of ESD in Lindblad type of
evolution, the following problems become prominent: estimating the exact time to
ESD for mixed entangled states; the factorization law for concurrence
or any other useful measure 
of entanglement for two-qubit mixed entangled state; a general scheme
for controlling ESD; a complete analysis of ESD in multipartite states
and the quasi-covariance of dynamical evolution.
One can have a full understanding of ESD only after all the problems
have been solved. 

\section*{Acknowledgment}
We would like to thank Subhashish Banerjee for valuable
discussions. SKG would like to thank Somdeb Ghose for his time.

\end{document}